\documentclass[fleqn,10pt]{wlscirep}   	

\usepackage{graphicx}				
\usepackage{graphicx}
\graphicspath{{fig/}}				
\usepackage{url}

\usepackage[skip=10pt,font=small,labelfont=bf,labelsep=period,justification=justified]{caption}
\usepackage[skip=5pt,font=footnotesize,format=hang,justification=centering,labelfont=normalfont]{subcaption}	
\DeclareCaptionLabelFormat{sublabel}{\bf#2.}
\DeclareCaptionLabelFormat{subreflabel}{\bf#2}
\usepackage{amssymb}
\usepackage{amsmath}
\usepackage{mathtools}

\newcommand\SetSymbol[1][]{
   \nonscript\,#1\vert \allowbreak \nonscript\,\mathopen{}}
   \providecommand\given{} 
\DeclarePairedDelimiterX\Set[1]{\lbrace}{\rbrace}%
 { \renewcommand\given{\SetSymbol[\delimsize]} #1 }
 
\DeclarePairedDelimiterXPP{\E}[1]{\operatorname{E}}{[}{]} {}{ \renewcommand\given{\SetSymbol[\delimsize]} #1 }
\DeclarePairedDelimiterXPP{\Var}[1]{\operatorname{Var}}{[}{]} {}{ \renewcommand\given{\SetSymbol[\delimsize]} #1 }
\DeclarePairedDelimiterXPP{\Prb}[1]{\operatorname{Pr}}{[}{]} {}{ \renewcommand\given{\SetSymbol[\delimsize]} #1 }
\renewcommand{\Pr}{\Prb}

\DeclareMathOperator{\kdelta}{\delta}
\DeclareMathOperator*{\argmax}{argmax}
\DeclarePairedDelimiterXPP{\function}[2]{\operatorname{#1}}{(}{)}{}{#2}
\DeclarePairedDelimiterXPP{\num}[1]{\#}{[}{]}{}{#1}

\newcommand{\mean}[1]{\langle#1\rangle}

\usepackage[capitalize]{cleveref}

\title{Multiresolution Consensus Clustering in Networks}
\author[1,*]{Lucas G. S. Jeub}
\author[2,3]{Olaf Sporns}
\author[1,3]{Santo Fortunato}

\affil[1]{School of Informatics, Computing and Engineering, Indiana University}
\affil[2]{Department of Psychological and Brain Sciences, Indiana University}
\affil[3]{Indiana University Network Science Institute (IUNI)}
\affil[*]{ljeub@iu.edu}
\date{Draft: \today}							

\begin{abstract}
Networks often exhibit structure at disparate scales. We propose a method for identifying community structure at different scales based on multiresolution modularity and consensus clustering. Our contribution consists of two parts. First, we propose a strategy for sampling the entire range of possible resolutions for the multiresolution modularity quality function. Our approach is directly based on the properties of modularity and, in particular, provides a natural way of avoiding the need to increase the resolution parameter by several orders of magnitude to break a few remaining small communities, necessitating the introduction of ad-hoc limits to the resolution range with standard sampling approaches. Second, we propose a hierarchical consensus clustering procedure, based on a modified modularity, that allows one to construct a hierarchical consensus structure given a set of input partitions. While here we are interested in its application to partitions sampled using multiresolution modularity, this consensus clustering procedure can be applied to the output of any clustering algorithm. As such, we see many potential applications of the individual parts of our multiresolution consensus clustering procedure in addition to using the procedure itself to identify hierarchical structure in networks.
\end{abstract}

\begin{document}
\maketitle
\flushbottom

\section*{Introduction}

Community detection \cite{Porter2009,Fortunato2010,Fortunato2016}, i.e., identifying groups of nodes that are densely connected internally and loosely connected to the rest of the system, is an important tool for analyzing the structure of networks. Community detection is an ill-defined problem where one typically identifies community structure as approximate solutions to difficult optimization problems (e.g., modularity~\cite{Newman2004a}, Infomap~\cite{Rosvall2008}, Stochastic Block Models~\cite{Karrer2011}). Heuristics for identifying good community structure based on these optimization problems are often stochastic in nature, returning different structures for different runs of the algorithm.  These different structures, even though they tend to be similar in quality based on the optimization criterion, can be rather different (see Good et al.\cite{Good2010} for examples in the context of modularity maximization). Furthermore, networks often exhibit structure at different scales and multiresolution community detection methods (e.g., based on Modularity \cite{Reichardt2006,Arenas2008}, Infomap \cite{Schaub2012}, and Stability \cite{Schaub2012a}), which identify structure at different scales based on a resolution parameter, have been developed to address this problem. For Stochastic Block Models, the number of blocks acts as a natural resolution parameter. 

As a result, when attempting to identify communities in a network, one is often confronted with the problem that one has a large number of potential partitions (as a result of different algorithms, multiple runs, and/or multiple values of a resolution parameter) and often no good way to select a single best partition. Consensus clustering (also known as ensemble clustering) \cite{Ghosh2011a,Vega-Pons2011} attempts to mitigate this problem by identifying common features of an ensemble of partitions. The combination of the partitions of the ensemble yields a consensus partition, which is representative of the ensemble, in that it is more similar, on average, to all network divisions than any one of them. We stress that consensus clustering is essentially a noise-reduction technique, that delivers robust results, but not necessarily a better solution of the problem. For instance, if we deal with partitions generated by the optimization of some objective function (e.g. modularity, as we do here), the best solution, according to the method, is the one corresponding to the largest value of the objective function, by definition. The issue of the relation between the solution provided by an algorithm and the ground truth(s) hidden in the data is much debated~\cite{peixoto15,schaub17,peel17,delmotte11}.
Consensus clustering approaches have been applied to the community detection problem previously \cite{Seifi2013,Sales-Pardo2007,Lancichinetti2012,Campigotto2013}. As in these existing approaches, we use the pairwise co-classification of nodes as the basis for our consensus clustering procedure. This approach is natural in the case of community detection, as co-classification defines a network and one can thus use existing community detection algorithms for the consensus clustering step. 

The key distinguishing feature of our consensus clustering procedure compared to these existing approaches is that we use a modified version of modularity with a null model based on the ensemble of partitions for the consensus clustering step. This null model allows us to assess statistical significance of co-classification. In particular, this means that our consensus clustering procedure does not identify communities in random networks (similar to what was observed by Campigotto et al.\cite{Campigotto2013} but without the need of setting an arbitrary threshold). We also exploit this property to obtain hierarchical consensus structures by recursively applying the procedure to clusters obtained at the previous step, stopping once a cluster has no significant sub-clusters. This recursive procedure results in much simpler (and thus easier to interpret) hierarchical structures than sweeping all possible threshold values\cite{Seifi2013} and avoids the computationally expensive sorting step needed by the procedure suggested by Sales-Pardo et al. \cite{Sales-Pardo2007}. 

Another procedure that is worth mentioning in this context is the belief propagation based modularity optimization procedure (ModBP) by Zhang and Moore\cite{Zhang2014}. This procedure only considers the consensus of an ensemble of partitions in a distributional sense and is not a consensus clustering procedure in the usual sense as it does not take sampled partitions as an input. As a result, it is not applicable in most situations where one would use a consensus clustering approach. However, this implicit approach is computationally very efficient. As a result ModBP can be used to analyze very large networks, which is not feasible using our proposed consensus clustering approach as well as previous approaches based on the coclassification matrix. ModBP also allows one to identify hierarchical community structure in networks using a recursive approach that is similar to our hierarchical consensus procedure. We compare ModBP and our hierarchical consensus procedure on synthetic benchmark networks with planted hierarchical structure.

The hierarchical aspect of our procedure is of particular interest for the application to ensembles of partitions obtained from multiresolution modularity that we focus on in this paper. In this application, one expects to see structure at different scales and a single consensus partition is unlikely to be particularly meaningful. In addition to the clustering method, the ensemble of input partitions itself is very important for consensus clustering to be successful. In the application to multiresolution modularity this means that we want our input partitions to cover all possible scales in the network as equally as possible. This is a non-trivial task as the possible values for the resolution parameter can often span several orders of magnitude and the sensitivity of obtained community structure on the value of the resolution parameter varies widely depending on the value of the resolution parameter itself. To address these problems, we propose a novel sampling strategy for the resolution parameter which we call {\it event sampling}. Event sampling directly exploits the behavior of the modularity quality function to provide good coverage of different scales in a network. We describe both the event sampling procedure and our Hierarchical Consensus procedure in Methods. Our Matlab implementation is available at \url{https://github.com/LJeub/HierarchicalConsensus}.

\section*{Results}

\subsection*{Comparison with Lancichinetti-Fortunato consensus clustering}

\begin{figure}[t!]
\begin{minipage}[b]{\linewidth}
\hfill\includegraphics{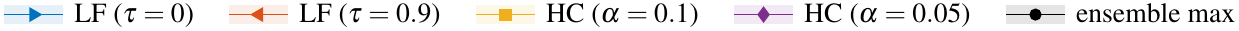}\hfill\hspace{0pt}
\end{minipage}\\

\hfill
\begin{subfigure}[b]{0.45\linewidth}
\includegraphics[width=\linewidth,height=0.6\linewidth]{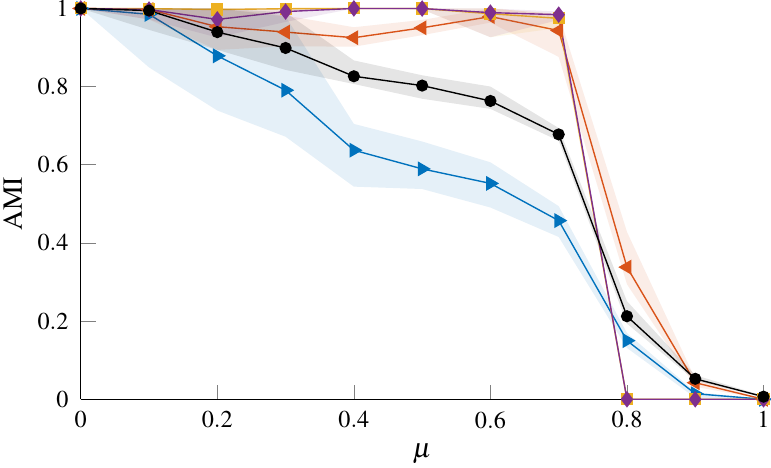}%
\caption{Fixed resolution ($\gamma=1$)\label{sfig:LF_comp_fixed}}
\end{subfigure}\hfill
\begin{subfigure}[b]{0.45\linewidth}
\includegraphics[width=\linewidth,height=0.6\linewidth]{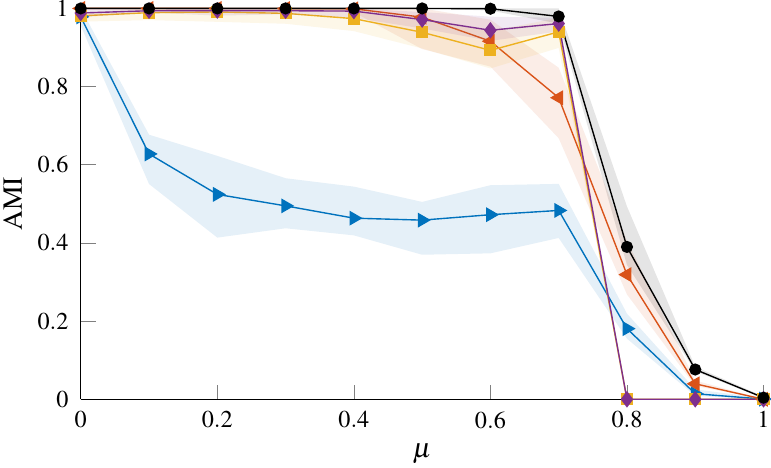}%
\caption{Multiresolution ensemble using event sampling\label{sfig:LF_comp_multi}}
\end{subfigure}\hfill\hspace{0pt}\\

\hfill
\begin{subfigure}[b]{0.45\linewidth}
\includegraphics[width=\linewidth,height=0.6\linewidth]{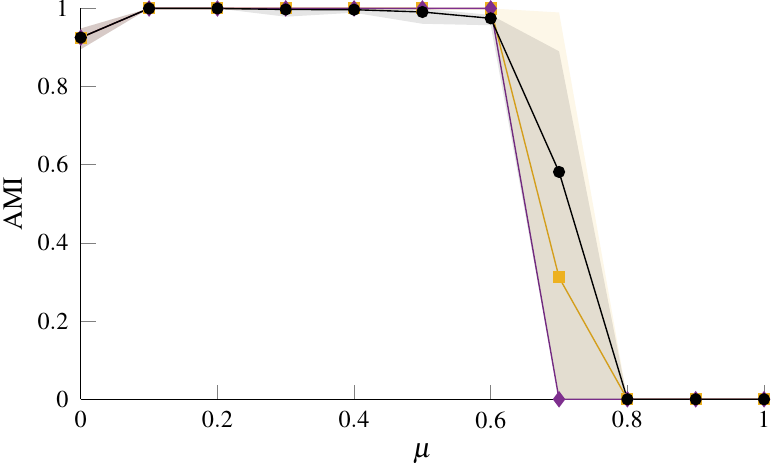}
\caption{Infomap ensemble\label{sfig:infomap}}
\end{subfigure}\hfill
\begin{subfigure}[b]{0.45\linewidth}
\includegraphics[width=\linewidth,height=0.6\linewidth]{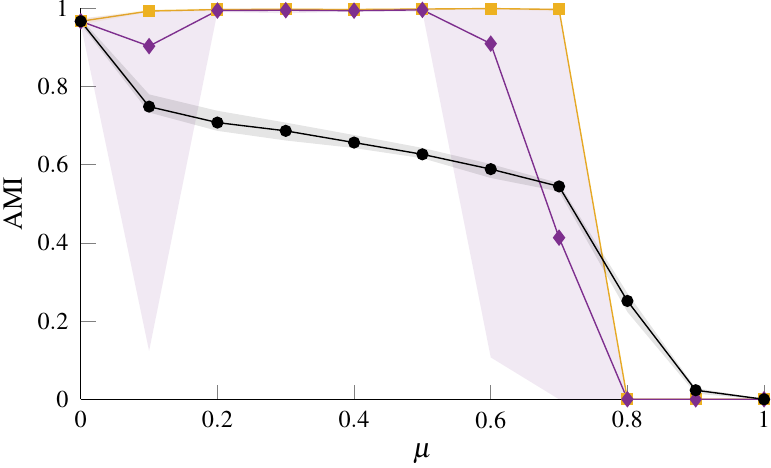}
\caption{Fixed resolution ($\gamma=1$) and 10000 nodes\label{sfig:LF_comp_10000}}
\end{subfigure}\hfill\hspace{0pt}

\caption{Comparison of the LF consensus procedure and our hierarchical consensus (HC) procedure on LFR benchmark networks. We use a sample of 10 networks for each value of the mixing parameter $\mu$. We compare the planted partition with the partitions identified using community detection using the adjusted mutual information (AMI), where the lines indicate the mean value and the shaded regions indicate the range of observed values for the sampled networks. The networks in panels \subref{sfig:LF_comp_fixed}--\subref{sfig:infomap} have 1000 nodes, power law degree distributions (mean degree $\mean{k}=20$, maximum degree $k_{\max}=50$, and exponent $\tau_1=2$), and  power law community size distributions (minimum size $c_{\min}=10$, maximum size $c_{\max}=50$, exponent $\tau_2=3$). The consensus partitions are based on ensembles with 250 partitions. In panel \subref{sfig:LF_comp_fixed}, the partitions are generated by optimizing modularity at a fixed resolution $\gamma=1$. In panel \subref{sfig:LF_comp_multi}, each partition is generated by optimizing modularity at a different value of $\gamma$ selected using the event sampling procedure (see methods). In panel \subref{sfig:infomap} we use Infomap\cite{Rosvall2008,infomap} to generate the partitions. The networks in panel \subref{sfig:LF_comp_10000} instead have 10000 nodes and a maximum community size $c_{\text{max}}=500$ and the original ensemble has 1000 partitions generated by maximizing modularity at $\gamma=1$. In each panel we include the results for the best partition of the original ensemble as a baseline (ensemble max). For the LF consensus procedure we chose two values of the threshold $\tau$ ($\tau=0$, which means that all entries of the consensus matrix are kept at each step, and $\tau=0.9$), while for our consensus procedure we selected two different values of the significance level $\alpha$ ($0.05$ and $0.1$). As a baseline, we include the performance of the partition in the initial ensemble that most closely resembled the planted partition (ensemble max).  \label{fig:LF_comp}}
\end{figure}

\begin{figure}[t!]
\begin{minipage}[b]{\linewidth}
\centering
\includegraphics{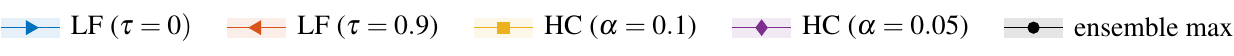}\\
\end{minipage}\\

\hfill
\begin{subfigure}[b]{0.45\linewidth}
\includegraphics[width=\linewidth,height=0.6\linewidth]{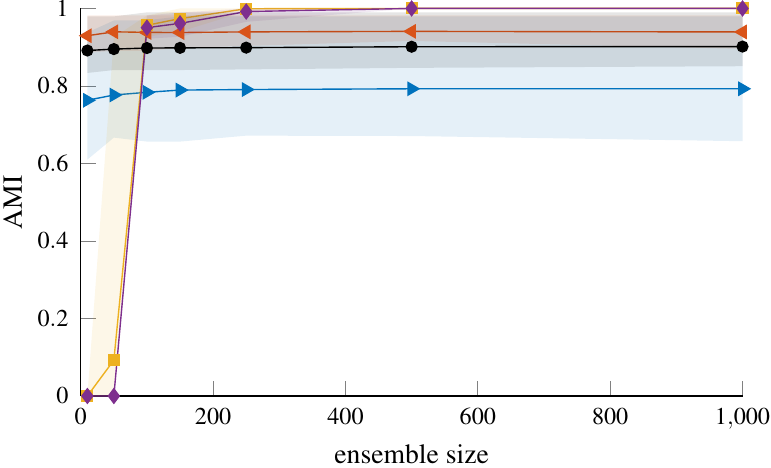}
\caption{$\mu=0.3$}
\end{subfigure}\hfill
\begin{subfigure}[b]{0.45\linewidth}
\includegraphics[width=\linewidth,height=0.6\linewidth]{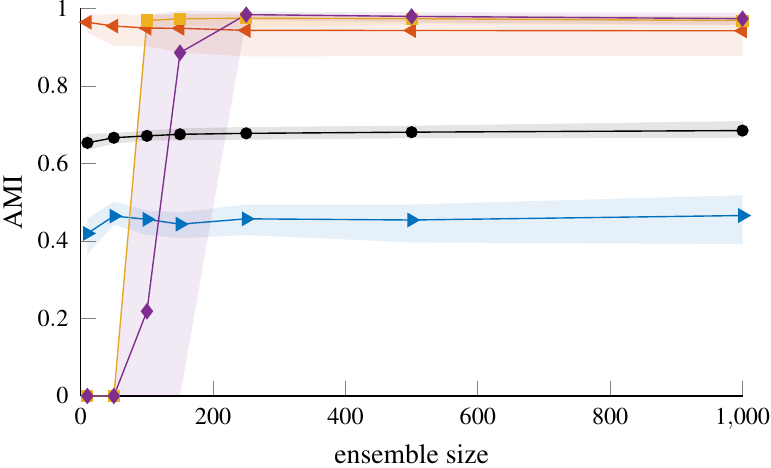}
\caption{$\mu=0.7$}
\end{subfigure}\hfill\hspace{0pt}\\

\hfill\begin{subfigure}[b]{0.45\linewidth}
\includegraphics[width=\linewidth,height=0.6\linewidth]{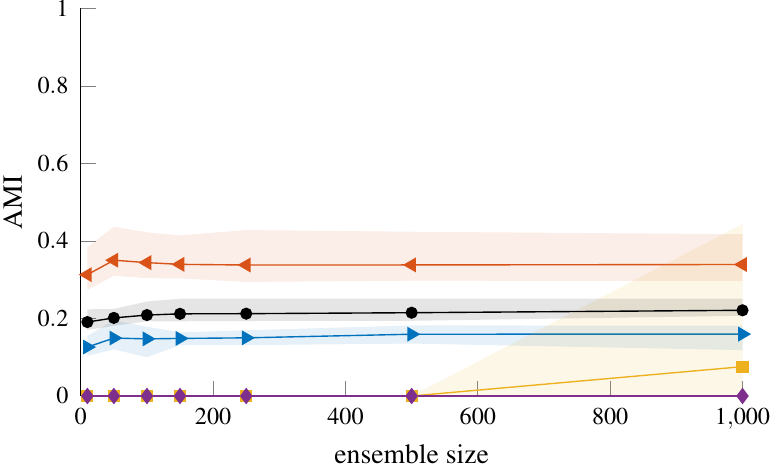}
\caption{$\mu=0.8$}
\end{subfigure}\hfill
\begin{subfigure}[b]{0.45\linewidth}
\includegraphics[width=\linewidth,height=0.6\linewidth]{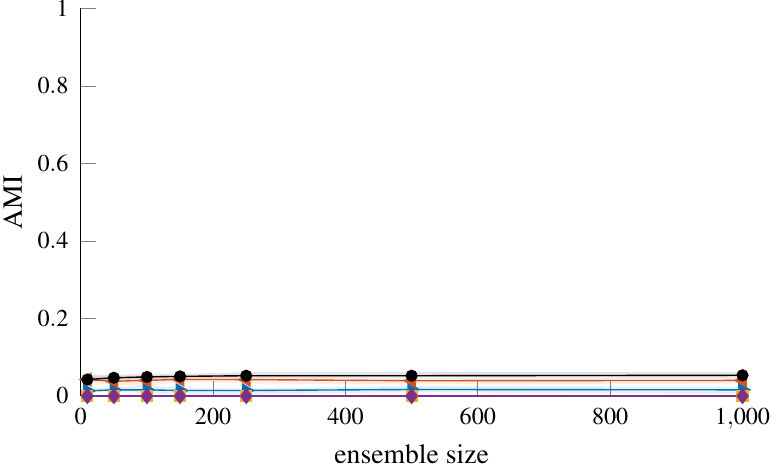}
\caption{$\mu=0.9$}
\end{subfigure}\hfill\hspace{0pt}

\caption{Effect of ensemble size on the performance of consensus clustering procedures. The networks are LFR benchmark graphs for four different values of the mixing parameter: $\mu=0.3$, $0.7$, $0.8$, $0.9$. The other parameters are the same as for the graphs used in Fig. 1. The LF consensus procedure is effectively independent of the size of the original ensemble, whereas our HC procedure needs a sufficiently large ensemble to identify the planted communities. \label{fig:LF_sample_comp}}
\end{figure}

In this section we compare the performance of our Hierarchical Consensus (HC) procedure with the iterative procedure proposed by Lancichinetti and Fortunato \cite{Lancichinetti2012} (the LF procedure). The LF procedure also uses the co-classification matrix to identify consensus clusters. It iteratively applies the same community detection algorithm that was used to obtain the original ensemble to a thresholded co-classification matrix for the ensemble of partitions generated at the previous iteration. The algorithm stops once all partitions in the ensemble are identical. At each iteration the co-classification matrix is thresholded, zeroing out small elements below a threshold $\tau\in[0,1]$.

We use LFR benchmark networks \cite{Lancichinetti2008,Lancichinetti2009a} to compare the performance of the two consensus clustering procedures. The parameter choices for the benchmark networks in \cref{fig:LF_comp} are the same as those originally used to test the LF procedure\cite{Lancichinetti2012}. These parameter choices produce networks with many small communities where modularity-based community detection methods (such as the iterated Louvain-like algorithm we use) run into the resolution limit of the measure \cite{Fortunato2007,Kumpula2007}, typically merging some of the communities in the output partitions. However, given that partitions merging different communities are of similar quality, one would expect that different partitions in an ensemble generated by a stochastic modularity based community detection algorithm may merge different small communities. Thus a consensus clustering algorithm may be able to recover the planted communities in this example. 

In \cref{sfig:LF_comp_fixed,sfig:LF_comp_multi,sfig:infomap}, we consider networks with 1000 nodes and compare different ways of constructing the initial ensemble of partitions used as input for the consensus clustering procedures. For \cref{sfig:LF_comp_fixed,sfig:LF_comp_multi}, we use a Louvain-like algorithm (see methods) to optimize multiresolution modularity. In \cref{sfig:LF_comp_fixed}, we use a fixed resolution of $\gamma=1$, which corresponds to the standard modularity\cite{Newman2004a} to sample the partitions. In \cref{sfig:LF_comp_multi}, we instead use the event sampling procedure (see methods) to select a  different value of $\gamma$ to sample each partition. In \cref{sfig:infomap}, we use Infomap\cite{Rosvall2008,infomap} to sample the partitions. In each case we generate ensembles with 250 partitions. In \cref{sfig:LF_comp_10000} we consider larger networks with 10000 nodes and generate ensembles with 1000 partitions by optimizing modularity at $\gamma=1$. 

 The similarity between the planted partition of the benchmark and the consensus partition(s) is estimated via the {\it adjusted mutual information}\footnote{Specifically the AMI${}_{\max}$ variant} (AMI) \cite{Vinh2010}. We choose AMI instead of the more popular {\it normalized mutual information} (NMI)~\cite{ana2003robust}, as the latter tends to overstate the similarity between partitions with many small clusters (see methods). We will be using AMI throughout this work. On the $x$-axis we have the {\it mixing parameter} $\mu$, that expresses how mixed the clusters are with each other ($\mu=0$ meaning that they are disjoint, $\mu=1$ that they have no internal edges). As a result, one expects the performance of community detection algorithms to decline as $\mu$ increases.

The results for the LF consensus procedure are very dependent on the value of the threshold $\tau$. In this example, we get the best results for large values of $\tau$. The best value of $\tau$ is situation dependent \cite{Lancichinetti2012} and it is not obvious how to choose $\tau$ in the absence of ground-truth information. While our recursive consensus clustering method also depends on a parameter $\alpha$, this has an intuitive interpretation as a statistical significance level and setting it to a reasonable value such as $\alpha=0.05$ usually provides good results. Furthermore, our method can detect when a network does not have significant community structure. In particular, we do not identify any communities in the LFR networks with $\mu\geq 0.8$ in \cref{sfig:LF_comp_fixed}, whereas both the LF consensus and the partitions in the original ensemble identify non-trivial communities that are essentially unrelated to the planted partition. The results for the HC procedure in \cref{fig:LF_comp} are for the finest level of the consensus hierarchy, i.e., for those clusters that do not have statistically significant sub-clusters. However, we note that in almost all cases our HC procedure identifies spurious intermediate hierarchical levels.

For any consensus clustering algorithm, the main goal is to return an output partition that is more representative of the underlying data than the input partitions. This baseline is represented by the performance of the partition in the initial ensemble that most closely resembles the planted partition (ensemble max) in \cref{fig:LF_comp}. As we can see from \cref{sfig:LF_comp_fixed,sfig:infomap,sfig:LF_comp_10000}, our HC procedure significantly outperforms this baseline in the case of fixed-resolution ensembles and recovers the planted partition almost perfectly even when the input partitions are of relatively low quality as is the case when using modularity optimization to generate the initial ensemble. Infomap already comes close to optimal performance on these networks, however, we still see a small improvement by using the HC procedure, especially when $\mu$ is large. 

The situation for the multiresolution ensemble in \cref{sfig:LF_comp_multi} is a bit more complicated. In this case, the best partition in the initial ensemble significantly outperforms the consensus partitions identified by either the LF procedure or our HC procedure, especially when $\mu$ is large. However, one should keep in mind that in the case of a multiresolution ensemble, the best partition corresponds to the partition at the optimal value of the resolution parameter. Identifying this optimal resolution a priori without knowledge of the planted partition is difficult (although recent work linking modularity optimization and the planted partition model\cite{Newman2016} suggest a potential way to do so). As such, the HC procedure comes close to identifying the optimal resolution.

A key parameter for the performance of the HC procedure is the size of the initial ensemble. In \cref{fig:LF_sample_comp} we compare the behavior of LF and HC consensus clustering as a function of the ensemble size (i.e. the number of partitions used to compute the co-classification matrix). As we can see, the performance of the LF procedure is essentially independent of the ensemble size. In contrast, our HC procedure needs a sufficiently large ensemble to identify any communities at all (where the minimum necessary ensemble size is larger for smaller $\alpha$) but outperforms the LF procedure for larger ensembles. This is particularly impressive given that we avoid the arbitrary thresholding of the co-classification matrix which is key to the performance of the LF procedure in this example. The minimum ensemble size for the HC method arises from the fact that for small ensembles even zero co-classification can no longer be considered significant evidence that two nodes are not in the same community. This results in a resolution limit that depends on the number of partitions in the initial ensemble and the desired significance level $\alpha$. Increasing the number of partitions in the initial ensemble allows one to identify smaller communities. Alternatively, increasing $\alpha$ also allows one to detect smaller communities given the same number of partitions in the initial ensemble at the cost of increasing $\alpha$ also increasing the risk of over-fitting. However, the results in \cref{fig:LF_comp,fig:LF_sample_comp} suggest that moderately increasing $\alpha$ (e.g., using $\alpha=0.1$ instead of $\alpha=0.05$) can improve the results and is a viable strategy if increasing the size of the initial ensemble is not possible.

\subsection*{Hierarchical Benchmark}

\begin{figure}[t!p]
\hfill\includegraphics{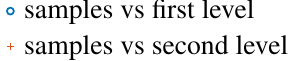}\includegraphics{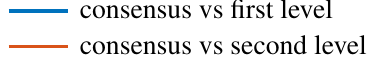}\hfill\hspace{0pt}\\
\smallskip
\hfill\begin{subfigure}{0.45\linewidth}
\includegraphics[width=\linewidth]{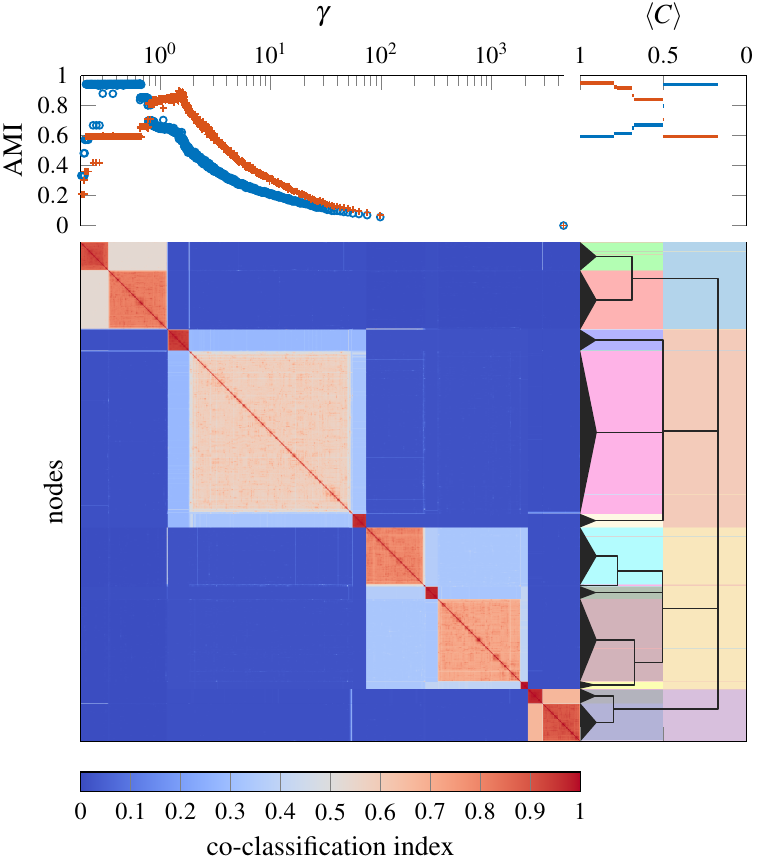}
\caption{Event sampling and local permutation model\label{sfig:bench-ill-ev-lp}}
\end{subfigure}\hfill
\begin{subfigure}{0.45\linewidth}
\includegraphics[width=\linewidth]{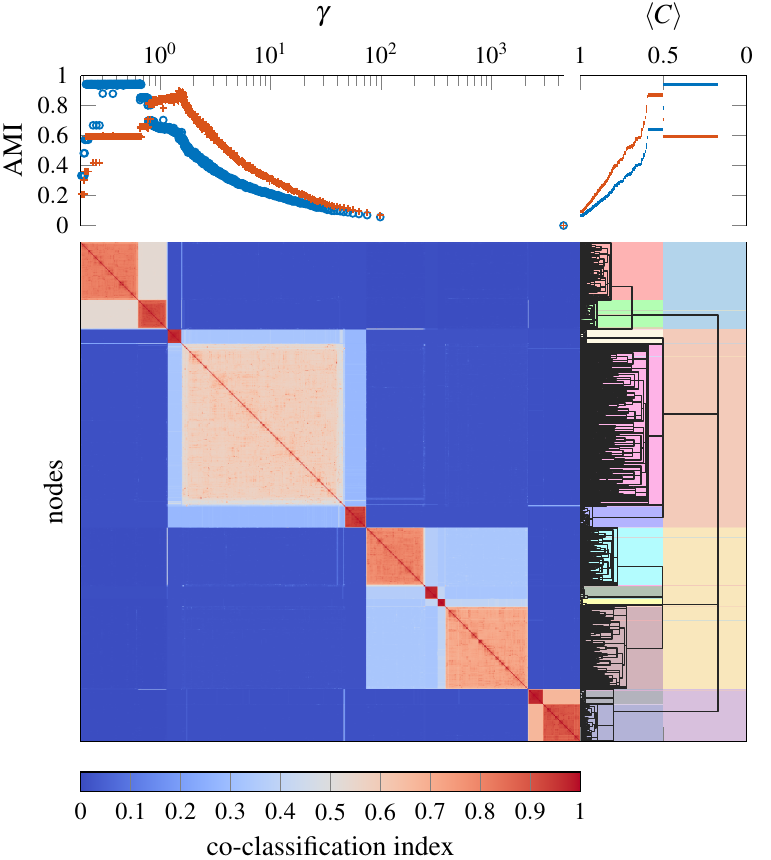}
\caption{Event sampling and permutation mode\label{sfig:bench-ill-ev-p}l}
\end{subfigure}\hfill
\smallskip

\hfill\begin{subfigure}{0.45\linewidth}
\includegraphics[width=\linewidth]{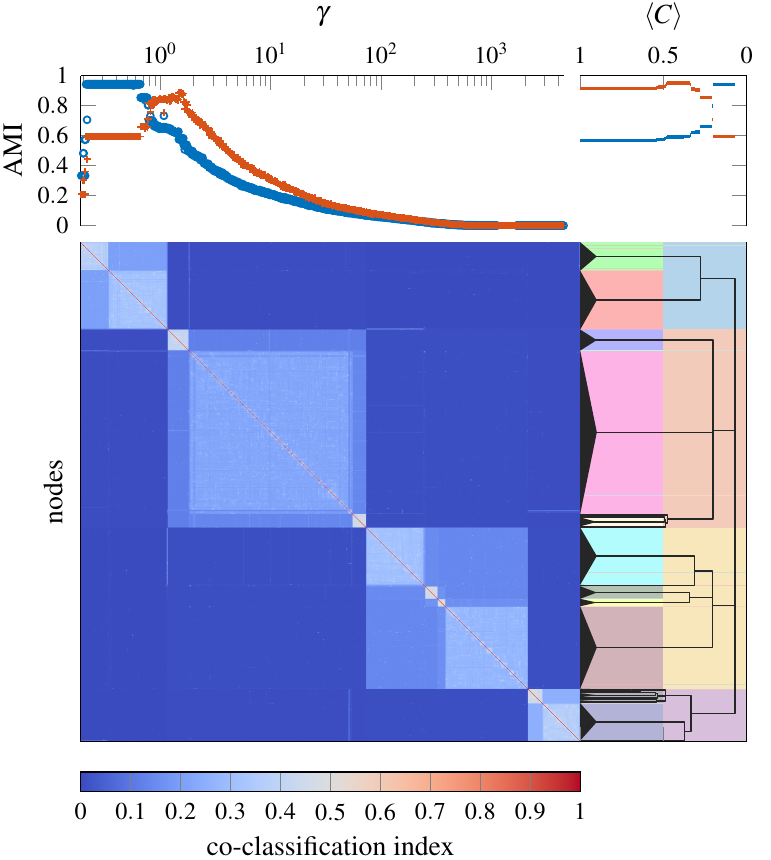}
\caption{Exponential sampling and local permutation model\label{sfig:bench-ill-ex-lp}}
\end{subfigure}\hfill
\begin{subfigure}{0.45\linewidth}
\includegraphics[width=\linewidth]{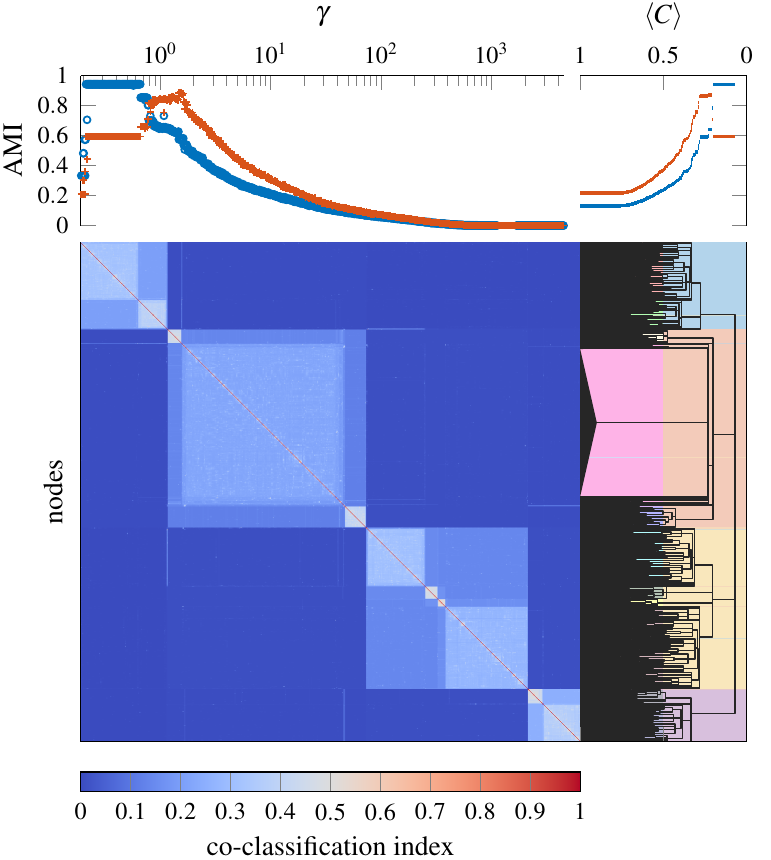}
\caption{Exponential sampling and permutation model\label{sfig:bench-ill-ex-p}}
\end{subfigure}\hfill\hspace{0pt}
\caption{Hierarchical benchmark example illustrating different sampling procedures and null models. The bottom part of each panel shows the co-classification matrix on the left and the planted ground truth (background colors) and consensus hierarchy on the right. The x-axis for the consensus hierarchy plot corresponds to the average value of the coclassifcation matrix $\mean{C}$ restricted to a given cluster (see Methods). The top part of each panel shows the comparison between (left) the ground truth partition and individual partitions for each value of $\gamma$  and (right) the ground truth partitions and the consensus hierarchy, estimated by the AMI. We compare the consensus hierarchy with the ground truth by considering cuts of the consensus hierarchy at each value of $\mean{C}$, merging clusters that split at a value of $\mean{C}$ greater than the currently considered value. Thus the value of the AMI for a given value of $\mean{C}$ corresponds to the partition obtained by drawing a vertical line through the consensus hierarchy at this value of $\mean{C}$.   \label{fig:bench-ill}}
\end{figure}

\begin{figure}[t!]
\begin{minipage}[b]{\linewidth}
\centering
\includegraphics{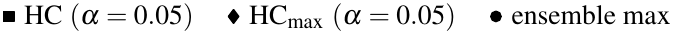}\\
\includegraphics{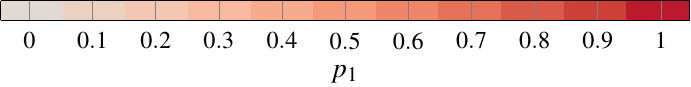}
\end{minipage}\\

\hfill\begin{subfigure}[b]{0.45\linewidth}
\includegraphics[width=\linewidth,height=0.6\linewidth]{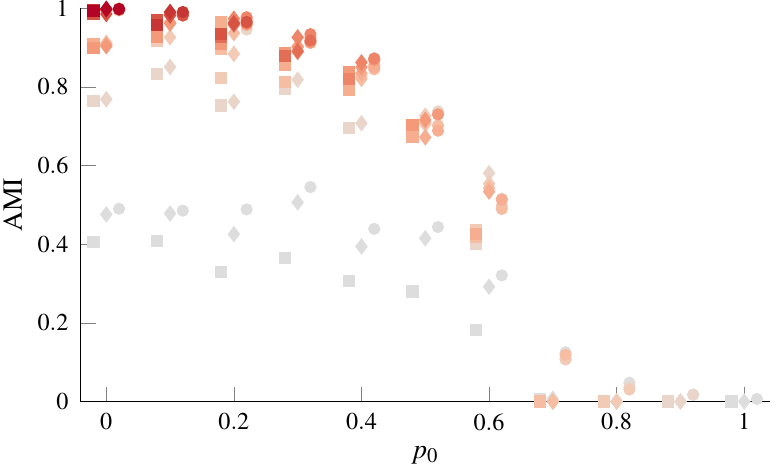}
\caption{First level of the planted hierarchy\\ (event sampling)\label{sfig:benchmark_ev_coarse}}
\end{subfigure}
\hfill
\begin{subfigure}[b]{0.45\linewidth}
\includegraphics[width=\linewidth,height=0.6\linewidth]{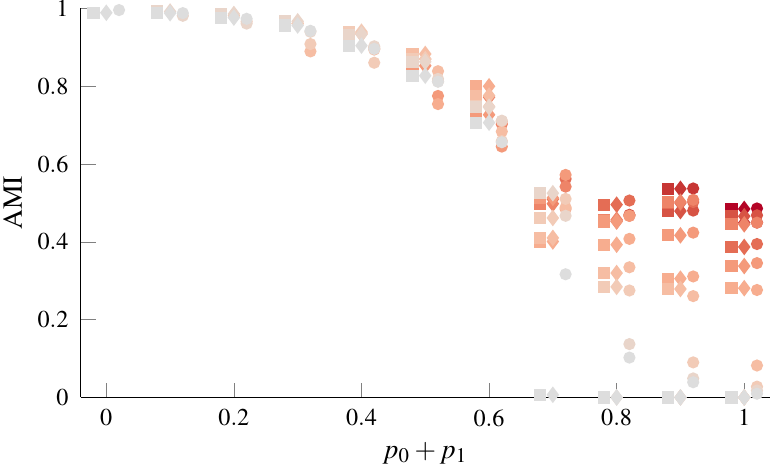}
\caption{Second level of the planted hierarchy\\ (event sampling)\label{sfig:benchmark_ev_fine}}
\end{subfigure}\hfill\vspace{0pt}\\

\hfill\begin{subfigure}[b]{0.45\linewidth}
\includegraphics[width=\linewidth,height=0.6\linewidth]{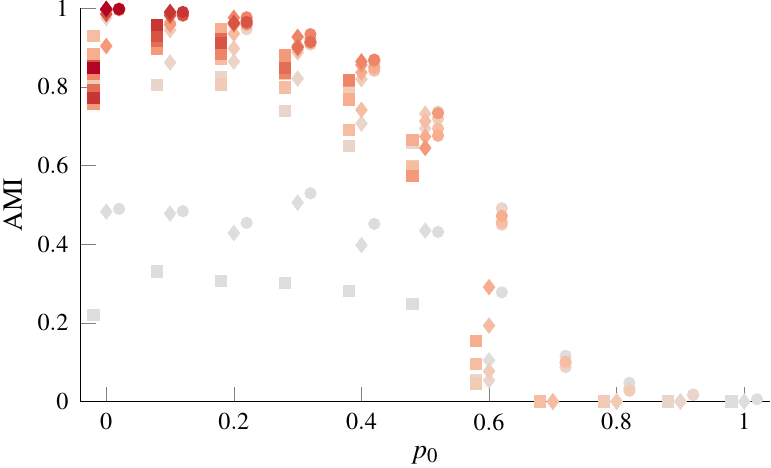}
\caption{First level of the planted hierarchy\\ (exponential sampling)\label{sfig:benchmark_log_coarse}}
\end{subfigure}
\hfill
\begin{subfigure}[b]{0.45\linewidth}
\includegraphics[width=\linewidth,height=0.6\linewidth]{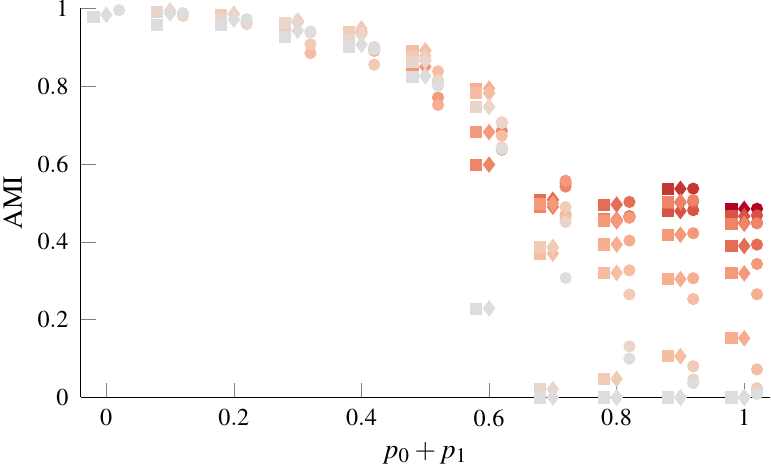}
\caption{Second level of the planted hierarchy\\ (exponential sampling)\label{sfig:benchmark_log_fine}}
\end{subfigure}\hfill\hspace{0pt}
\caption{Performance of hierarchical consensus clustering on two-level hierarchical benchmark networks with 1000 nodes. The parameters for the benchmark are chosen such that $p_i\in \{0,0.1,\ldots,1\}$ with the constraint that $p_0+p_1+p_2=1$. The results shown in this figure are averaged over 10 independent realizations of the benchmark for each combination of the parameters. We generate ensembles of 250 partitions for these networks using multiresolution modularity with event sampling and exponential sampling of the $\gamma$-range. In panels \subref{sfig:benchmark_ev_coarse} and \subref{sfig:benchmark_log_coarse}, we consider the first level planted partition of the benchmark and compare it to the coarsest partition identified by the consensus hierarchy (HC) and the best partition identified by cuts of the consensus hierarchy (HC$_{\text{max}}$). In panels \subref{sfig:benchmark_ev_fine} and \subref{sfig:benchmark_log_fine} we instead consider the second level planted partition and compare it to the finest partition identified by the consensus hierarchy (HC) and the best partition identified by cuts of the consensus hierarchy (HC$_{\text{max}}$). We include the performance of the best individual partition in the original ensembles (ensemble max) as a baseline. Data points are color-coded based on the value of $p_1$ (i.e., the fraction of edges constrained to lie within the first-level partition) of the corresponding networks. Note that the first-level partition should be considered as noise when trying to identify the second level partition which is the reason for the different choice of x-axis in \subref{sfig:benchmark_ev_fine}, \subref{sfig:benchmark_log_fine}. \label{fig:benchmark_alpha0.95local}}
\end{figure}

\begin{figure}
\begin{minipage}[b]{\linewidth}
\centering
\includegraphics{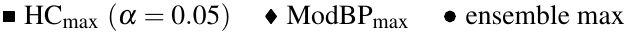}\\
\includegraphics{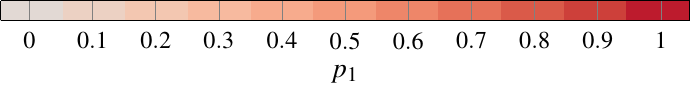}
\end{minipage}\\

\hfill\begin{subfigure}[b]{0.45\linewidth}
\includegraphics[width=\linewidth,height=0.6\linewidth]{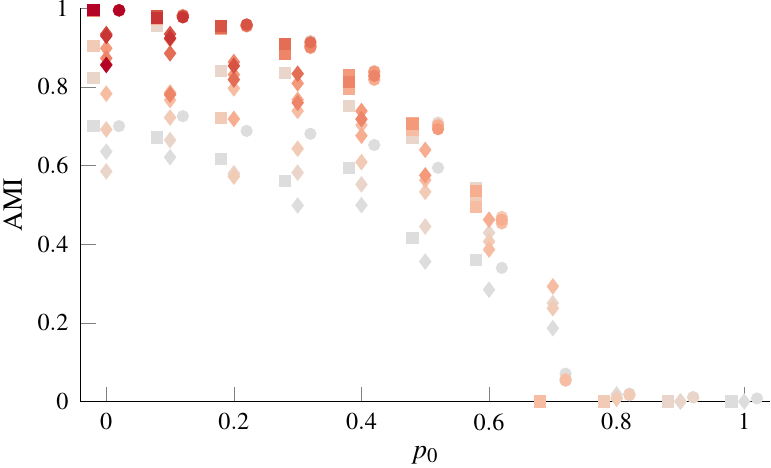}
\caption{First level of the planted hierarchy}
\end{subfigure}
\hfill
\begin{subfigure}[b]{0.45\linewidth}
\includegraphics[width=\linewidth,height=0.6\linewidth]{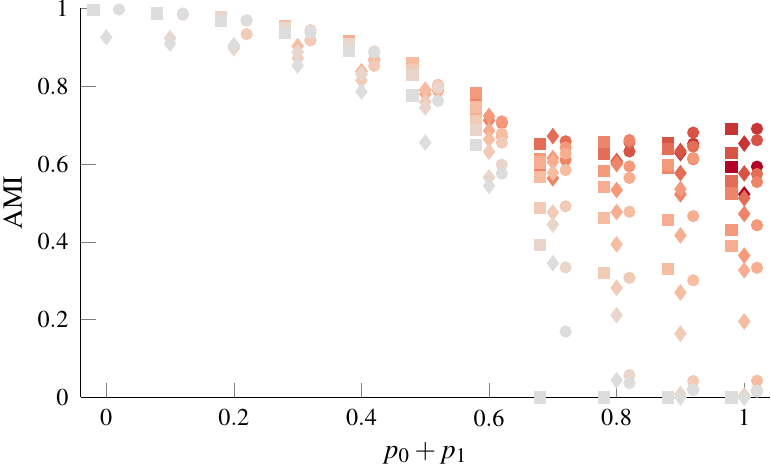}
\caption{Second level of the planted hierarchy}
\end{subfigure}\hfill\hspace{0pt}
\caption{Performance of hierarchical consensus clustering (HC) and belief-propagation based modularity optimization (ModBP) on two-level hierarchical benchmark networks with 10000 nodes. Except for the number of nodes, the procedure for generating the networks is the same as in \cref{fig:benchmark_alpha0.95local}. For the HC procedure we generate initial ensembles with 1000 partitions using multiresolution modularity with event sampling. We show the results for the best partition identified by the hierarchical consensus hierarchy (HC$_\text{max}$), the best partition identified by the belief-propagation hierarchy (ModBP$_\text{max}$), and the best partition in the original ensemble (ensemble max).\label{fig:benchmark_10000_modbp}} 
\end{figure}

We illustrate the effectiveness of our consensus clustering procedure to extract hierarchical structure on the example of artificial benchmark networks with two-level hierarchical community structure. The benchmark networks are based on a degree-corrected stochastic block model\cite{Karrer2011} with power-law degree distribution, with exponent $2$, minimum degree $5$ and maximum degree 70. Each of the networks has 1000 nodes and a 2-level hierarchical community structure, where a fraction $p_0$ of edges is allocated at random, only based on node degrees, a fraction $p_1$ of edges is constrained to lie within groups of the first level partition $\vec{g}_1$ (the one with larger clusters), and a fraction $p_2$ of edges is constraint to lie within groups of the second level partition $\vec{g}_2$ (the one with smaller clusters). Note that $p_0+p_1+p_2=1$. The community assignments are generated as follows:

\begin{enumerate}
\item All nodes are in a single community in the base level partition.
\item Subsequent levels are generated by splitting communities from the previous level:
\begin{enumerate}
\item Sample the number $c$ of new communities from a Poisson distribution with mean $4$ and minimum cutoff $2$ (to ensure each community is split).
\item Sample assignment probabilities for each of the new communities from a symmetric Dirichlet distribution with concentration parameter $\sigma=1.5$. The probabilities determine the size of the clusters.
\item Assign to each node a community label based on these probabilities.
\end{enumerate}
\end{enumerate}

We illustrate the results of different sampling procedures for $\gamma$ and different null models for the consensus clustering in \cref{fig:bench-ill} on a single instantiation of this benchmark with $\vec{p}=(0.2,0.2,0.6)$. These parameters correspond to a network with clear hierarchical community structure. The event sampling ensemble and exponential sampling ensemble contain 1000 partitions each (where each partition is sampled using a different value of $\gamma$) and we use a significance level of $\alpha=0.05$ to identify the consensus hierarchy. We see that using event sampling and the local permutation model (\cref{sfig:bench-ill-ev-lp}), we can recover the hierarchical structure almost perfectly.  In particular, we recover the first level of the planted hierarchy as accurately as any single partition in the ensemble and the second level more accurately than any single partition. As we can see in \cref{sfig:bench-ill-ev-p}, when using the permutation model we can identify the underlying structure initially but fail to detect the point at which we should stop splitting clusters further. As we can see from \cref{sfig:bench-ill-ex-lp,sfig:bench-ill-ex-p}, exponential sampling yields an ensemble that includes many partitions carrying essentially no information about the structure of the network. However, the identified consensus structure remains mostly similar to that obtained using event sampling with a few additional errors. While we only show a single illustrative example in \cref{fig:bench-ill}, we observe qualitatively very similar behavior across a wide range of networks including the real networks discussed later. 

The hierarchical consensus procedure only extracts a structural hierarchical tree which indicates which small clusters should be merged to form larger-scale clusters. It does not provide any information about the order in which merges in different branches of the tree should be performed to construct partitions that are representative of different scales in a network. However, the coclassification matrix contains additional information about the clusters which we can exploit. In particular, high coclassification of nodes within a cluster indicates a strong cluster whereas low coclassification indicates a weak cluster. Based on this idea, one can construct different measures for the strength of a cluster. This allows one to visualize the hierarchy produced by the HC procedure as a dendrogram, similar to the result of traditional hierarchical clustering procedures\cite{Hastie2009}. One can cut this dendrogram to construct partitions of the network where one only performs merges that result in clusters that are stronger than a given threshold. For the experiments in this paper we use the mean value of the coclassification matrix restricted to the newly created cluster $\mean{C}$ as our measure of cluster strength. However, as the HC procedure does not consider the strength of a cluster when constructing the hierarchy, there is no guarantee that a given measure is consistent with the resulting hierarchy and one may have to consider different measures in some applications.

We analyze the performance of our consensus clustering procedure on these benchmark networks in more detail in  \cref{fig:benchmark_alpha0.95local}. We consider networks with 1000 nodes and different combinations of the parameters $p_0$, $p_1$ and $p_2$. Comparing the results when using event sampling or exponential sampling for the $\gamma$-values used to generate the initial ensemble of partitions, we see that using event sampling improves our ability to identify the planted structure by a small amount. Using event sampling also improves the amount of noise we can tolerate. In the case of event sampling, we can still identify meaningful structure when $p_0=0.6$ whereas we fail to identify any structure in the networks using exponential sampling at this point. 

When assessing the performance of consensus clustering algorithms in the multiresolution case, one should keep in mind that the best individual partition is typically a much tougher baseline in the case of a multiresolution ensemble than in the case of a fixed resolution ensemble (compare \cref{sfig:LF_comp_fixed,sfig:LF_comp_multi}).
For the first-level partition (\cref{sfig:benchmark_ev_coarse,sfig:benchmark_log_coarse}), our HC procedure results in a consensus partition that tends to be slightly worse than the best individual partition in the initial ensemble. Furthermore, we see a significant improvement in performance when considering the best cut of the consensus hierarchy rather than the coarsest partition identified by the consensus hierarchy. This indicates the presence of spurious coarse clusters in the consensus hierarchy that merge several ground-truth clusters. For the second-level partition, the consensus partition tends to perform better than the best individual partition, provided $p_0+p_1<0.6$. When $p_0+p_1>0.6$, the finest level of the consensus hierarchy tends to more closely resemble the first-level planted partition rather than the second-level partition. Results for the best cut of the consensus hierarchy and the finest partition identified by the consensus hierarchy are essentially identical, indicating that our HC procedure successfully identifies when a community should not be split further. 

In \cref{fig:benchmark_10000_modbp}, we consider the effect of increasing the network size on our ability to extract the hierarchical structure. We construct networks in the same way as for \cref{fig:benchmark_alpha0.95local} but with 10000 nodes instead. As in the case of the LFR networks in \cref{sfig:LF_comp_10000}, we need to increase the size of the initial ensemble to achieve comparable results. However, given a sufficiently large initial ensemble, the HC procedure recovers the first level of the planted hierarchy about as well as the best partition in the initial ensemble. When compared with the second level of the planted hierarchy, the HC procedure slightly outperforms the best partition of the initial ensemble. We also compare the performance with belief-propagation based modularity optimization (ModBP)\cite{Zhang2014}, which can also identify hierarchical structure. Comparing the results for ModBP and HC on these networks, we see a small but consistent advantage of HC over ModBP for both levels of the planted hierarchy. However, one should keep in mind that the improved performance of HC over ModBP comes at a significant computational cost. ModBP is highly scalable and can be applied to very large networks, whereas HC at least in its current implementation is limited by its $O(n^2)$ memory requirements and more suited to the detailed exploration of smaller networks.

 Overall, our HC procedure can successfully identify hierarchical structure based on a multiresolution ensemble of partitions. One limitation of our HC procedure is that it tends to be conservative when splitting communities. As a result, we often see spurious intermediate levels in the consensus hierarchy which are the result of merging ground-truth clusters.

\subsection*{Real Networks}

\newcommand{\consensussubfigure}[4][0.45\linewidth]{
\begin{subfigure}[b]{#1}
\centering
\footnotesize
\includegraphics{{#2_consensus_ev_alpha#3_sample}.pdf}\hspace{0pt}\includegraphics{{#2_consensus_ev_alpha#3_cons}.pdf}\\
\includegraphics[width=\linewidth]{{#2_consensus_ev_alpha#3}.pdf}
\caption{#4}
\end{subfigure}
}

\begin{figure}[tp!]
\hfill\consensussubfigure{karate}{0.05_new}{Zachary Karate Club \cite{Zachary1977}}\hfill
\consensussubfigure{football}{0.05_new}{College Football\cite{Girvan2002,Evans2010}}\hfill\hspace{0pt}\\

\hfill\consensussubfigure{polblogs}{0.05_new}{Political Blogs\cite{Adamic2005}}\hfill
\consensussubfigure{polbooks}{0.05_new}{Political Books\cite{Krebs}}\hfill\hspace{0pt}
\caption{Consensus hierarchical community structure identified using a multiresolution event sampling ensemble with 1000 partitions and the local permutation model at significance level $\alpha=0.05$. The background colors under the consensus hierarchy indicate the reference ground-truth partition that we use to evaluate our results. \label{fig:gt_ex}}
\end{figure}

\begin{figure}
\centering
\includegraphics{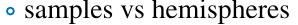}\includegraphics{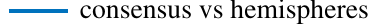}\\
\includegraphics[width=0.7\linewidth]{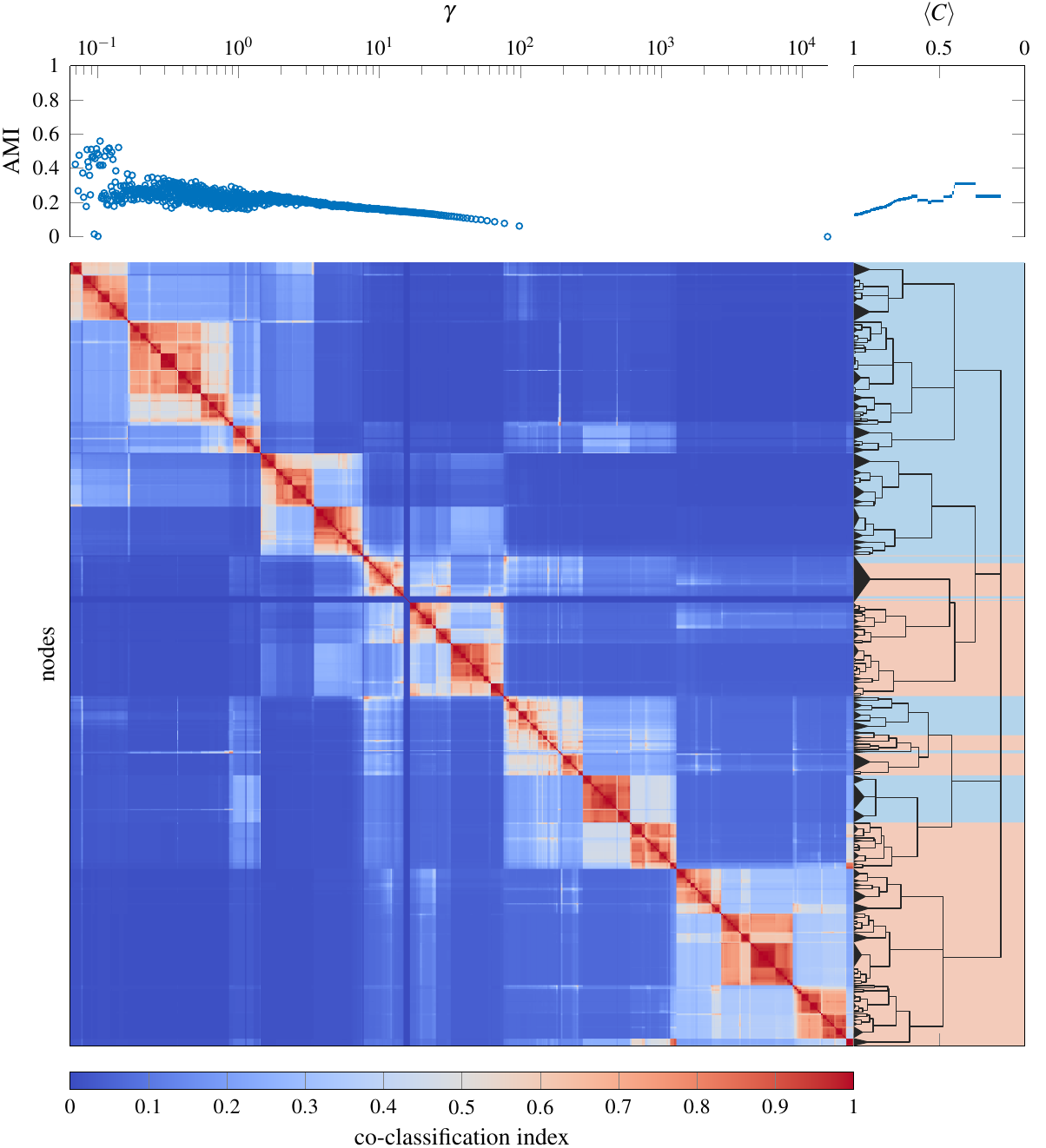}

\caption{Consensus hierarchical community structure for the Human Structural Brain Network\cite{Hagmann2008} identified using a multiresolution event sampling ensemble with 1000 partitions at significance level $\alpha=0.05$. The background colors under the consensus hierarchy indicate the division of the brain into the two hemispheres. \label{fig:brain}}

\end{figure}

\begin{figure}[t!]
\centering
\includegraphics{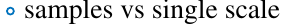}\includegraphics{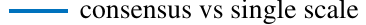}\\
\includegraphics[width=\linewidth]{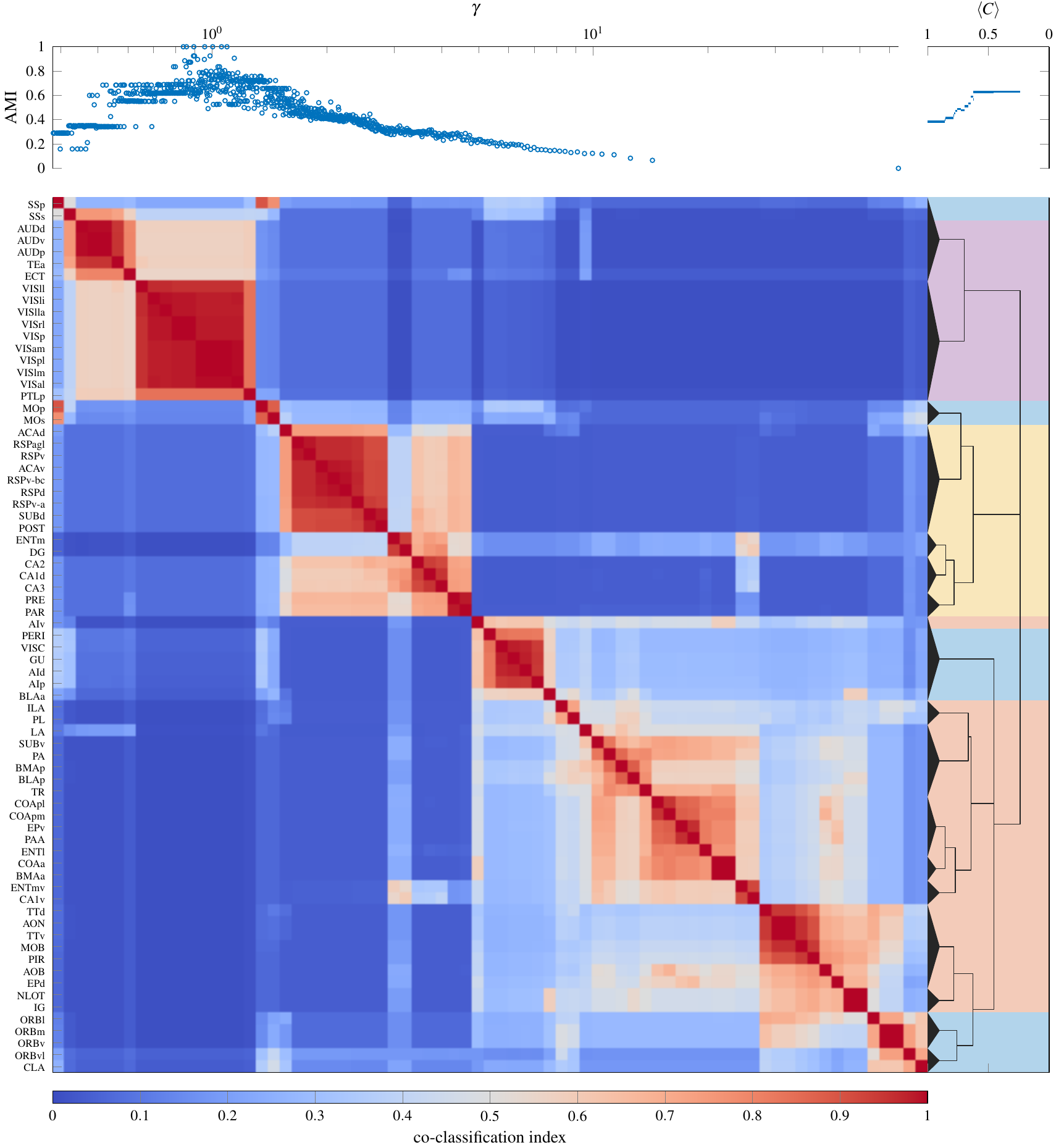}

\caption{Consensus hierarchical community strucutre for the Rat Structural Brain Network\cite{Bota2015} identified using a multiresolution event sampling ensemble with 1000 partitions at significance level $\alpha=0.05$. The background colors under the consensus hierarchy indicate the single scale partition extracted previously\cite{Bota2015}. Detailed descriptions for the abbreviated area names are available in the original publication\cite{Bota2015}. \label{fig:rat-brain}}
\end{figure}

In addition to artificial benchmark networks, we also use our consensus clustering procedure to identify community structure in networks with some notion of ground truth community structure (see \cref{fig:gt_ex}), a human structural brain network (see \cref{fig:brain}), and a rat structural brain network (see \cref{fig:rat-brain}). The networks we consider are:
\begin{itemize}
\item Zachary Karate Club \cite{Zachary1977}: Social network between the 34 members of a Karate club with ground truth partition given by the split of the original club into two new clubs that occurred during the original study.
\item American College Football (corrected) \cite{Girvan2002,Evans2010}: Network of American football games between the 115 Division IA colleges during regular season (Fall 2000) with ground truth partition given by conference membership.
\item Political Blogs \cite{Adamic2005}: Network of hyperlinks between 1224 blogs on US politics with ground truth partition given by political orientation (conservative or liberal).
\item Political Books \cite{Krebs}: Co-purchasing network of 105 books about US politics with ground truth partition given by political orientation (conservative, neutral, or liberal).
\item Human Structural Brain Network \cite{Hagmann2008}: Undirected, weighted connections derived from diffusion imaging and tractography between 998 regions of interest of the human cerebral cortex (averaged over 5 participants). Ground truth is lacking for this network as the community structure of brain networks is generally unknown. In place of ground truth partitions, we provide hemispheric ordering as a reference. 
\item Rat Structural Brain Network \cite{Bota2015}: Directed, weighted connections derived from histological tract tracing data between 73 nodes in the rat cerebral cortex. As with the human network, we do not have a notion of ground truth for this network. Instead, we compare our results to a previously published \cite{Bota2015} single scale partition.
\end{itemize}

As we can see from \cref{fig:gt_ex}, the best partition based on our consensus hierarchy, when compared to the ground truth, performs as well  as the best partition for any single value of $\gamma$ in our ensemble. At the same time, our consensus hierarchy provides a much simpler and easier to interpret description of the underlying structure than the entire ensemble of partitions. With the exception of the College Football network, the notion of ground truth corresponds most closely to the coarsest identified partition for these networks. In all these networks (even in the case of College Football where our results suggest that some conferences are organized into two subgroups) we identify structure that is smaller-scale than the ground truth partition. 

For the consensus partitions of the structural brain networks in \cref{fig:brain,fig:rat-brain} we observe a deeply nested hierarchical community structure, confirming previous studies that have suggested the existence of multiscale communities in brain anatomical and functional systems \cite{Betzel2016}. In the case of the human brain network in \cref{fig:brain}, we find that communities fall onto spatially contiguous parts of the cerebral cortex, in line with the prevailing idea that structural modules are spatially compact. In addition, we observe a strong association between pairwise co-classification and physical distance (estimated as the Euclidean distance) across all node pairs (Spearman's $\rho = -0.654$). The rat multiscale analysis recovers most of the modules that were extracted earlier \cite{Bota2015} based on a single-scale analysis, while also revealing new associations that were previously missed and are functionally meaningful, for example the association of somatosensory areas (SSp, SSs) with other primary sensory (auditory and visual) areas. In addition to retrieving larger modules similar to those described earlier, the multiscale consensus approach provides a detailed hierarchical view of how modules at different scales are arranged, a central objective for analyses of modular brain networks \cite{Sporns2016}.

In \cref{fig:gt_ex,fig:brain,fig:rat-brain} we only report results for a significance level of $\alpha=0.05$ and ensembles with 1000 partitions. In general, increasing the number of partitions in the ensemble allows one to potentially resolve finer-scale structure and smaller values of $\alpha$ reduce the risk of identifying spurious structure.  However, we observe very similar structure at a significance level of $\alpha=0.01$ in these examples, suggesting that our results are reasonably robust to these parameter choices. 

\section*{Discussion}

Our goal for this paper was to address two key issues that frequently arise in the context of community detection in networks. One often obtains many different community structure solutions (e.g., as a result of different runs of a stochastic algorithm) and one often expects networks to have meaningful community structure at different scales. To address these problems, we develop a hierarchical consensus clustering algorithm that can identify hierarchical community structure in networks based on an ensemble of input partitions. Our hierarchical clustering algorithm eliminates the need to select an arbitrary threshold for the co-classification matrix that is necessary with existing consensus clustering procedures \cite{Lancichinetti2012,Seifi2013} to achieve good results. This is important, as the value of the threshold strongly influences the results of these algorithms and the optimal value of the threshold is situation dependent and usually cannot be inferred from the available data. In artificial examples, where one can optimize the threshold to maximize recovery of the planted community structure, our hierarchical algorithm is competitive with the consensus clustering algorithm of \cite{Lancichinetti2012} for this optimized value of the threshold. Furthermore, our hierarchical consensus clustering algorithm does not identify communities in random networks that do not have community structure. 

To identify hierarchical structure in networks, we use multiresolution modularity \cite{Reichardt2006} to sample the initial ensemble. We  suggest a sampling strategy for the resolution parameter that ensures good coverage of all scales and should also be useful in other contexts (e.g., identifying stable partitions \cite{Arenas2008}, mesoscopic response functions \cite{Onnela2012}). We then use our hierarchical consensus algorithm to identify a consensus hierarchical community structure which automatically identifies relevant scales in the network without the need of selecting a particular value of the resolution parameter. Furthermore, the consensus hierarchy can combine features from partitions obtained at different resolutions, thus potentially avoiding the resolution limit problems inherent in the use of modularity. In both artificial and real-world examples with a notion of ground truth communities, the best partition identified by the consensus hierarchy is usually comparable to the best partition of the original ensemble. However, the consensus hierarchy typically only consists of a few levels of nested community structures and is thus much easier to interpret and provides a much simpler representation of network community structure than the original ensemble of partitions.

While we only considered multiresolution modularity to generate the initial ensemble, one could, in principle, use any multiresolution method instead. However, our event-sampling technique is specific to quality functions that can be decomposed as a sum over vertex pairs. Exponential and linear sampling, while not a good fit for multiresolution modularity, are general and may be good choices for other techniques.

\section*{Methods}
\subsection*{Multiresolution Modularity}

We use cluster assignment vectors $\vec{g}=[\num*{\vec{g}}]^n$ to denote a partition of a network into clusters, where $g_i$ denotes the cluster assignment of node $i$ (given as an integer between $1$ and $\num*{\vec{g}}$, where $\num*{\vec{g}}$ is the number of clusters). We use $\num*{{\vec{g}=c}}$ to denote the number of nodes in cluster $c$. Here we use the Reichart and Bornholdt \cite{Reichardt2006} version of the modularity quality function with a resolution parameter, 
\begin{equation} \label{eq:modularity}
Q(\vec{g}, \gamma)=\sum_{i,j=1}^{n} \left(A_{ij}-\gamma P_{ij}\right)\kdelta(g_i,g_j)\, ,
\end{equation}
where $A$ is the adjacency matrix, $P$ is the expected adjacency matrix under a null model, and $\gamma$ is a resolution parameter which can be tuned to influence the sizes of the clusters. A typical choice \cite{Newman2004a} for the null model (based on the assumption of fixed node degrees with edges otherwise placed at random) is 
\begin{equation}
P_{ij}=\frac{k_i k_j}{2m}\, , \qquad k_i=\sum_{j} A_{ij}\, , \qquad 2m=\sum_i k_i\, .
\end{equation}
We use this null model for clustering the original networks. For the consensus clustering step we use a different null model which we introduce later. We use the iterated version of the GenLouvain algorithm \cite{Jeub2011} to optimize $Q$ for all experiments in this paper.

\subsection*{Sampling Strategies for $\gamma$}

\begin{figure}
\hfill\includegraphics{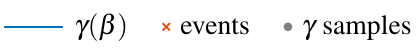}\hfill\hspace{0pt}\\

\begin{subfigure}[b]{0.32\linewidth}
\includegraphics[width=\linewidth,height=0.75\linewidth]{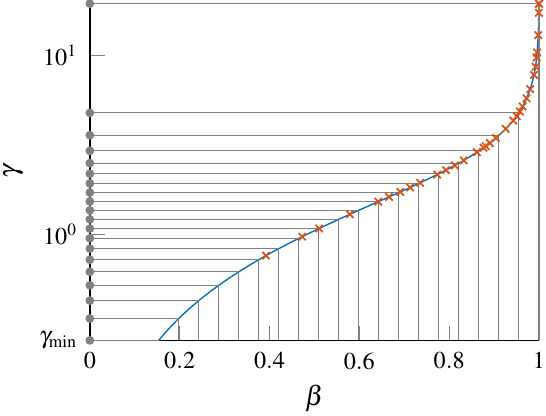}
\caption{Zachary Karate Club}
\end{subfigure}\hfill
\begin{subfigure}[b]{0.32\linewidth}
\includegraphics[width=\linewidth,height=0.75\linewidth]{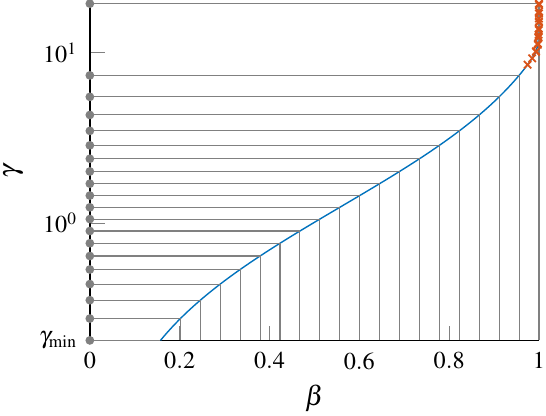}
\caption{College Football}
\end{subfigure}\hfill
\begin{subfigure}[b]{0.32\linewidth}
\includegraphics[width=\linewidth,height=0.75\linewidth]{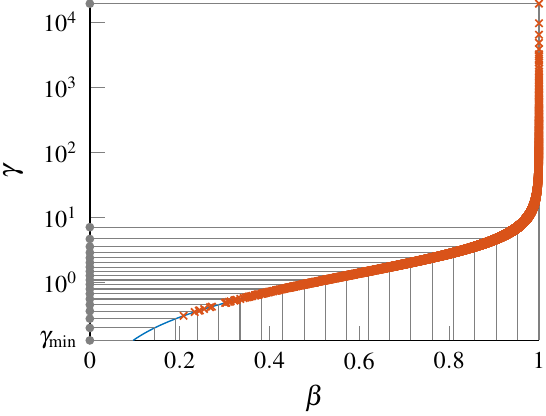}
\caption{Political Blogs}
\end{subfigure}\\

\begin{subfigure}[b]{0.32\linewidth}
\includegraphics[width=\linewidth,height=0.75\linewidth]{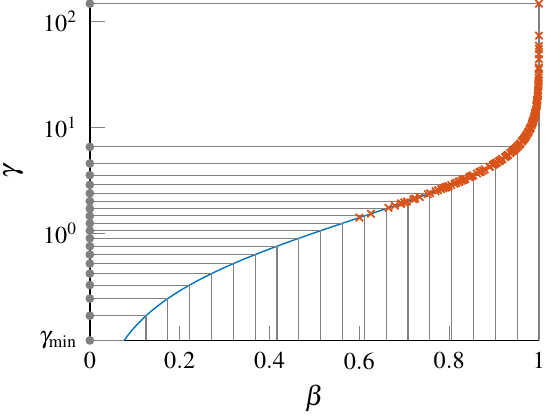}
\caption{Political Books}
\end{subfigure}\hfill
\begin{subfigure}[b]{0.32\linewidth}
\includegraphics[width=\linewidth,height=0.75\linewidth]{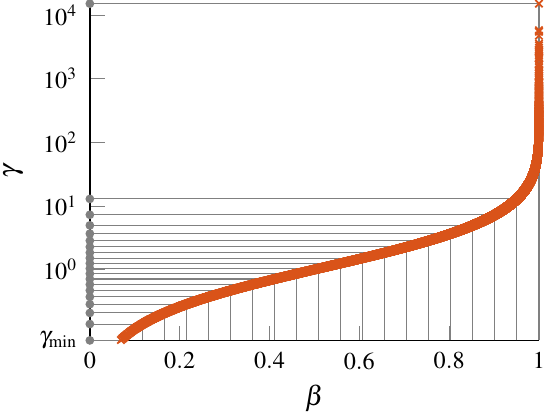}
\caption{Human Structural Brain}
\end{subfigure}\hfill
\begin{subfigure}[b]{0.32\linewidth}
\includegraphics[width=\linewidth,height=0.75\linewidth]{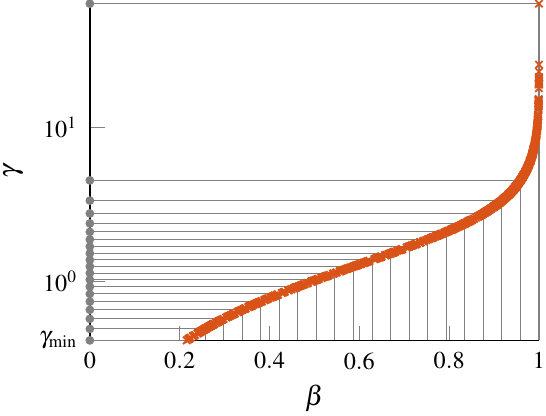}
\caption{Rat Structural Brain}
\end{subfigure}

\caption{Illustration of the event sampling procedure for generating $\gamma$ samples. We take equally spaced samples (20 in this figure) in $\beta$ and map them to the corresponding $\gamma$ values using \cref{eq:gamma_beta}. Note how this procedure automatically avoids excessive sampling of large $\gamma$ values as the fraction of antiferromagnetic contributions is approaching 1 in this regime and increases very slowly as we increase $\gamma$ further. \label{fig:eventsampling}} 
\end{figure}

For multiresolution modularity (\cref{eq:modularity}) one can define a meaningful range $[\gamma_{\mathrm{min}},\gamma_{\mathrm{max}}]$ of $\gamma$ values that covers all possible resolutions, where 
\begin{equation}
\begin{aligned}
\gamma_{\mathrm{min}}&=\max \Set{\gamma \given \num*{\vec{g}_{\mathrm{max}}(\gamma)}=1}\\
\gamma_{\mathrm{max}}&=\min \Set{\gamma \given \num*{\vec{g}_{\mathrm{max}}(\gamma)}=n}\\
\vec{g}_{\mathrm{max}}(\gamma)&=\argmax_{\vec{g}} Q(\vec{g},\gamma)
\end{aligned}
\end{equation}
The upper bound $\gamma_{\mathrm{max}}$ can be computed exactly by noting that $\gamma_{\mathrm{max}}$ is the smallest value of $\gamma$ such that $A_{ij} - \gamma P_{ij} \leq 0$ for all $i$ and $j$. However, $\gamma_{\mathrm{min}}$ has to be determined numerically. To estimate $\gamma_{\mathrm{min}}$ we use that modularity is a linear function of $\gamma$ for a fixed partition. This means that we can directly compute the minimum value of $\gamma$ for which a given partition is better than the trivial partition where all nodes are in the same community. This is effectively a special case of the CHAMP algorithm\cite{Weir2017}. We can use this observation to iteratively estimate $\gamma_{\mathrm{min}}$. The iterative algorithm proceeds by first estimating $\gamma_{\mathrm{min}}$ using a small sample of partitions at $\gamma=1$. We then sample a new set of partitions using $\gamma=\gamma_{\mathrm{min}}-\epsilon$ (to ensure the previous partitions are strictly non-optimal) and use the new sample to update $\gamma_{\mathrm{min}}$. We repeat this process until the new sample consists only of the trivial partition.

We want to sample from this range of $\gamma$ values in a way that ensures that we give equal coverage to different scales in the network. Obvious sampling strategies would be linear sampling and exponential sampling. For linear sampling one would use equally spaced values of $\gamma$ between $\gamma_{\mathrm{min}}$ and $\gamma_{\mathrm{max}}$ and for exponential sampling one would use values that are equally spaced on a logarithmic scale. However, these sampling strategies do not work well in practice when considering the entire range of gamma values. In particular, for many networks there is a large range of gamma values where the network is almost completely fragmented into singletons with a few small communities remaining that take extremely large values of $\gamma$ to split apart. Especially for linear sampling (and to a lesser extend also for exponential sampling) many of the sampled partitions will be from this regime and thus carry very little information about the network structure.

\subsubsection*{Event Sampling}

To avoid this issue we introduce a new sampling strategy that we dub event sampling. The event sampling strategy is inspired by an idea used by Onnela et al.\cite{Onnela2012} to produce mesoscopic response functions that are comparable across different networks. As in Onnela et al.\cite{Onnela2012} we split the contributions of node pairs to the modularity into ferromagnetic interactions $E^+(\gamma)=\Set{(i,j) \given i\neq j, A_{ij} - \gamma P_{ij} > 0}$ and antiferromagnetic interactions $E^-(\gamma)=\Set{(i,j) \given i \neq j, A_{ij}-\gamma P_{ij} < 0}$. Onnela et al.\cite{Onnela2012} used the fraction of antiferromagnetic interactions, however we found this not to give a good coverage of the different scales in a network. Noting that the behavior of modularity depends on the relative magnitudes of ferromagnetic and antiferromagnetic interactions, as a single strong interaction can compensate for many weak interactions, we instead propose to use the relative magnitude of antiferromagnetic interactions
\begin{equation}
\beta(\gamma) = \frac{\sum_{(i,j)\in E^-(\gamma)} |A_{ij}-\gamma P_{ij}|}{\sum_{(i,j),i\neq j} |A_{ij}-\gamma P_{ij}|}
\end{equation}
as a measure of scales. Note that $\beta(\gamma)=0$ for $\gamma \leq 0$ and $\beta(\gamma)=1$ for $\gamma\geq \gamma_{\max}$ and is monotonically increasing for $0 \leq \gamma \leq \gamma_{\max}$.

To sample $\gamma$, we invert the relationship between $\gamma$ and $\beta$ 
\begin{equation}
\gamma(\beta)=\frac{\sum_{(i,j) \in E^-(\beta)} A_{ij} + \beta \left(\sum_{(i,j) \in E^+(\beta)}A_{ij} - \sum_{(i,j) \in E^-(\beta)}A_{ij} \right)}{\sum_{(i,j) \in E^-(\beta)} P_{ij} + \beta \left(\sum_{(i,j) \in E^+(\beta)}P_{ij} - \sum_{(i,j) \in E^-(\beta)}P_{ij} \right)} \label{eq:gamma_beta}
\end{equation}
and sample $\gamma(\beta)$ at equally spaced values of $\beta$ between $\beta_{\mathrm{min}}$ and $\beta_{\mathrm{max}}$, where $\beta_{\mathrm{min}}=\beta(\gamma_{\mathrm{min}})$ and $\beta_{\mathrm{max}}=\beta(\gamma_{\mathrm{max}})$ (see \cref{fig:eventsampling}). To compute $E^+(\beta)$ and $E^-(\beta)$, note that these change only at a discrete set of values (the ``events'' in \cref{fig:eventsampling}) which are straightforward to identify.

\subsection*{Consensus Clustering}

The idea behind consensus clustering is to combine multiple partitions for the same network to obtain a (ideally more meaningful) consensus partition for the network.   We use an approach that is based on the co-classification matrix $C(\mathbf{g})$ for a set of partitions $\mathbf{g}=\{\vec{g}{(t)}\}_{t=1}^{|\mathbf{g}|}$, defined as
\begin{equation}
\label{eq:co-classification}
C_{ij}(\mathbf{g}) = \frac{1}{|\mathbf{g}|} \sum_{t=1}^{|\mathbf{g}|} \kdelta(g_i{(t)},g_j{(t)})\, .
\end{equation}
The co-classification matrix defines a new network. As a result one can in principle apply any network clustering method that can handle weighted networks to the co-classification matrix to obtain a consensus partition \cite{Lancichinetti2012}.  However, the co-classifcation matrix has a peculiar structure that we would like to exploit. One of the strengths of the modularity quality function is that it is often straightforward to input prior knowledge about the structure of a network by using an appropriate null model to compute $P$.

\subsubsection*{Null Models for Consensus Clustering}

For the consensus clustering problem we can define appropriate null models based on the input partitions. Assuming that under the null model different partitions are independently sampled from different distributions, we can write the entries $C_{ij}^0$ of the co-classification matrix under the null model $\mathbf{g}^0$ as a normalized sum of independent Bernoulli-distributed random variables
\begin{equation}
	C_{ij}^0 = \frac{1}{|\mathbf{g}^0|}\sum_{t=1}^{|\mathbf{g}^0|}  C_{ij}^0(t) \; ,\qquad C_{ij}^0(t) \sim \function*{Bernoulli}{p^0_{ij}(t)}\; , \qquad p_{ij}^0(t)=\Pr*{g_i^0(t) = g_j^0(t)} \; .
\end{equation}
This means that $C_{ij}^0$ follows a rescaled Poisson-Binomial distribution. The full distribution of $C_{ij}^0$ is complicated to characterize, however, we can easily compute its mean and variance:
\begin{align}
\label{eq:mean}
\E*{C_{ij}^0} &= \frac{1}{|\mathbf{g}^0|} \sum_{t=1}^{|\mathbf{g}^0|} \E*{C_{ij}^0(t)} = \frac{1}{|\mathbf{g}^0|}  \sum_{t=1}^{|\mathbf{g}^0|} p_{ij}^0(t)=\mu_{ij}\\
\Var{C_{ij}^0} &= \frac{1}{|\mathbf{g}^0|^2} \sum_{t=1}^{|\mathbf{g}^0|} \Var*{C_{ij}^0(t)}= \frac{1}{|\mathbf{g}^0|^2} \sum_{t=1}^{|\mathbf{g}^0|} p_{ij}^0(t) - \left(p_{ij}^0(t)\right)^2=\sigma^2_{ij} \;.
\end{align}
From \cref{eq:mean} we get that using $P_{ij}=\mu_{ij}$ in \cref{eq:modularity} is a sensible choice for consensus clustering. Furthermore, we can assess significance of co-clustering by estimating the distribution of $C_{ij}^0$, e.g., using its asymptotic normal approximation $C_{ij}^0\sim \function{N}{\mu_{ij},\sigma_{ij}}$, or by pseudorandom sampling of the Bernoulli trials. We use the asymptotic normal approximation in our experiments as it is computationally more efficient and produces similar results to pseudo-random sampling in practice.

To determine the $p_{ij}^0(t)$, we need to make further assumptions about what it means to have a random partition. Different assumptions lead to different null models \cite{Gates2017} which we describe below. Note that the case $i=j$ is trivial as $\kdelta(g_i(t) , g_i(t))=1$ for any partition and hence $p_{ii}^0(t)=1$ irrespective of our null-assumptions. For simplicity, we thus assume that $i\neq j$ from now on.

\subsubsection*{Permutation Model}
The permutation model is perhaps the most common null model for ensembles of partitions. Its null assumptions are that for each sampled partition the number and sizes of clusters are fixed but nodes are otherwise assigned to clusters at random. Under these assumptions one obtains the following formula for $p_{ij}^{\text{perm}}(t)$ given a sample of partitions $\mathbf{g}$:
\begin{equation}
\label{eq:perm}
\begin{aligned}
p_{ij}^{\text{perm}}(t)&=\sum_{c=1}^{\num*{\vec{g}(t)}}  \Pr*{{g^0_i(t)=c},{g^0_j(t)=c}}=\sum_{c=1}^{\num*{\vec{g}(t)}} \Pr*{g^0_i(t)=c}\Pr*{g^0_j(t)=c \given g^0_i(t)=c}\\
&=\sum_{c=1}^{\num*{\vec{g}(t)}}\frac{\num*{{\vec{g}(t)=c}}}{n}\frac{\num*{{\vec{g}(t)=c}}-1}{n-1}
\end{aligned}
\end{equation}
Note that under this null model all nodes are equivalent and hence $C_{ij}$ has the same distribution for any pair of nodes such that $i\neq j$. It is hence a constant null model and thus in a sense resolution free \cite{Traag2011}. While the simplicity of this model is appealing, we find that it does not work well in practice, in particular in the recursive application. It often results in splitting of individual nodes or small groups of nodes from clusters in the artificial networks and real networks (see \cref{fig:bench-ill} for an example). This problem arises because the presence of small groups has only a very small effect on the expected co-classifcation of nodes under this model as most nodes are assigned to the large group (and hence co-classified). This means that a single partition in an ensemble where a node is in a singleton group can often be considered significant evidence that a node is not part of a cluster under this model.

\subsubsection*{Local Permutation Model}

To avoid this problem we introduce the local permutation model. As in the permutation model, we assume that the sizes and number of clusters are fixed. Additionally, when computing $p_{ij}^{\text{lperm}}(t)$, we assume that the community assignment of node $i$ is fixed and only that of node $j$ is random. Hence, given a sample of partitions $\mathbf{g}$, we have
\begin{equation}
p_{ij}^{\text{lperm}}(t) = \Pr*{g_j^0(t)=g_i(t) \given g_i^0(t) = g_i(t)} = \frac{\num*{\vec{g}(t) = g_i(t)}-1}{n-1}
\end{equation}
This model seems to give a good compromise between identifying clusters that should be split and not splitting clusters that should not be split. Note that in general $p_{ij}^{\text{lperm}}\neq p_{ji}^{\text{lperm}}$ and we address how we deal with this technicality later.

\subsubsection*{Other Models}

We also considered the other two null models suggested by Gates et al.\cite{Gates2017}, namely the fixed number of clusters model and the uniform random partition model. However, these do not perform well in this application. In particular, the fixed number of clusters model can be either too conservative or too aggressive, depending on the relative sizes of clusters, whereas the uniform random partition model is so conservative that one would almost never identify any communities at all. We would expect however that in situations where one has some additional information about the structure of the ensemble of partitions, taking those into account in the null model may improve results.

\subsection*{Consensus Modularity}

We use a modularity-like quality function for the consensus clustering step with a null matrix that is based on statistical significance of co-classification under the local permutation model, i.e.,
\begin{equation}
\label{eq:cons_mod}
Q_C(\vec{g},\alpha) = \sum_{i,j} \left( C_{ij} - P_{ij}^{\text{lperm}}(\alpha)\right) \delta(g_i,g_j)\, ,
\end{equation}
where
\begin{equation}
P_{ij}^{\text{lperm}}(\alpha) = p \text{ such that }  \function*{\max}{\Pr*{C_{ij}^\text{lperm}\leq p}, \Pr*{C_{ji}^\text{lperm}\leq p}} = \alpha\, .
\end{equation}
This choice of null matrix means that the only negative contributions to the sum in \cref{eq:cons_mod} are from pairs of nodes that are statistically significantly (at significance level $\alpha$) less frequently co-classified than could be explained by the local permutation model where the community assignment of one of the nodes remains fixed. 

To obtain a consensus partition we use the iterative procedure suggested by Lancichinetti and Fortunato\cite{Lancichinetti2012}. First, we obtain a new ensemble of partitions by optimizing \cref{eq:cons_mod} using iterated GenLouvain\cite{Jeub2011}. We then compute a new co-classification matrix for the new ensemble and repeat the procedure until the co-classification matrix is binary, i.e., until all partitions in the ensemble are identical (typically after one or two iterations). We do not use any thresholding on the co-classification matrices as the GenLouvain procedure is designed to handle full matrices.

\subsection*{Hierarchical Consensus Procedure}

A single partition is often not a good consensus summary for a sample of partitions. In the case of multiresolution modularity, this may be because of multiple meaningful scales in a network or because of the resolution limit problems of the modularity quality function \cite{Fortunato2007}. We propose a recursive strategy for generating a hierarchical cluster tree to extract more information.

Starting from a partition where all nodes are in the same community, we apply the following procedure:
\begin{enumerate}
\item For each cluster of the partition apply modularity based consensus clustering at a given significance level (so that we do not split clusters that could result at random) restricted to the nodes that are in the cluster
\item repeat the procedure for each newly generated cluster
\end{enumerate}
This procedure stops once there are no more clusters that can be split into subclusters at the given significance level and results in a tree of clusters. We refer to this procedure as the hierarchical consensus (HC) procedure.

\subsubsection*{Computational Complexity}

The computational complexity of the hierarchical consensus procedure is largely determined by the need to store and manipulate the co-classification matrix which is typically dense. This results in an $O(n^2)$ memory requirement and also limits the efficiency of the community detection algorithms used to cluster the co-classification matrix. The computational complexity of the hierarchical consensus procedure itself is largely determined by the the time needed to identify the first level of the hierarchy. The computational complexity of the Louvain-like community detection algorithm we use to cluster the co-classification matrix is approximately $O(m)$ where $m$ is the number of positive entries of the consensus modularity matrix (see \cref{eq:cons_mod}). This results in an overall complexity of $O(lm)$ for the hierarchical consensus procedure, where $l$ is the number of partitions in the initial ensemble. Given that the hierarchical consensus procedure is trivially parallelizable, as one can sample partitions independently, scalability is typically memory-limited in practice. 

It is in principle possible to reduce the memory requirements and thus potentially improve scalability by noting that the coclassification matrix can be written as 
\[C=\tfrac{1}{l} G G^T,\]
where $G$ is the node-cluster adjacency matrix for the initial ensemble. Note that $G$ is sparse with exactly $nl$ non-zero entries, one for each node-partition pair. The current version of the GenLouvain code does not exploit this type of structure efficiently.

\subsection*{Modularity Optimization Algorithm}

We use the iterated variant of the GenLouvain code\cite{Jeub2011} with weighted random moves. This algorithm is similar to the original Louvain algorithm\cite{Blondel2008} and uses the same two-phase structure. The algorithm starts with all nodes in their own community. During the first phase, the algorithm tries to move single nodes between communities, choosing between moves that increase modularity at random with probability proportional to the resulting increase in modularity. When there are no more single-node moves that increase modularity, the identified communities are aggregated to form new supernodes in phase two. Phase one and two are repeated with the supernodes until no further increase in modularity is identified. The algorithm is then restarted repeatedly using the new communities as the initial partition until no further increase in modularity is identified. This restarting of the algorithm can often yield significant improvements in modularity.

\subsection*{Adjusted Mutual Information}

Normalized variants of the mutual information $I(\vec{g},\vec{h})$ between two partitions $\vec{g}$ and $\vec{h}$ are frequently used to compare partitions. The mutual information measures the amount of shared information between the two partitions and is given by
\begin{equation*}
    I(\vec{g},\vec{h})=\sum_{c=1}^{\num*{g}} \sum_{d=1}^{\num*{h}} \Prb*{g_i=c,h_i=d} \log \frac{\Prb*{g_i=c,h_i=d}}{\Prb*{g_i=c}\Prb{h_i=d}}
\end{equation*}
where $i$ is a random node. The mutual information can be normalized in different ways, e.g., by noting that $I(\vec{g},\vec{h})\leq\max \{ H(\vec{g}),H(\vec{h})\}$, where 
\begin{equation*}
H(\vec{g})= -\sum_{c=1}^{\num*{g}} \Prb*{g_i=c} \log \Prb*{g_i=c}
\end{equation*}
is the entropy of partition $\vec{g}$. This yields a normalized mutual information
\[
NMI_{\text{max}}(\vec{g},\vec{h})=\frac{I(\vec{g},\vec{h})}{\max\{H(\vec{g}),H(\vec{h})\}}.
\]
However, this measure and other variants of normalized mutual information are badly behaved for partitions with many clusters\cite{Vinh2010}. The adjusted mutual information\cite{Vinh2010}
\[
AMI_{\text{max}}= \frac{I(\vec{g},\vec{h})-\E{I(\vec{g},\vec{h})}}{\max \{H(\vec{g}),H(\vec{h})\} - \E{I(\vec{g},\vec{h})}}
\]
mitigates this problem by correcting the measure based on the expected value $\E{I(\vec{g},\vec{h})}$ of the mutual information under the permutation model.

\bibliography{consensus_bib}

\section*{Acknowledgements}
Computational resources used by this research were supported in part by the National Science Foundation under Grant No. CNS-0521433, in part by Lilly Endowment, Inc., through its support for the Indiana University Pervasive Technology Institute, and in part by the Indiana METACyt Initiative. The Indiana METACyt Initiative at IU was also supported in part by Lilly Endowment, Inc.

\section*{Author Contributions}
LJ, OS and SF devised the study. LJ performed the numerical experiments, analyzed the data and prepared the figures. LJ and SF wrote the main manuscript text. All authors reviewed the manuscript.

\section*{Additional Information}
{\bf Competing financial interests:} The authors declare no competing financial interests.

\end{document}